%% file: main.tex
\documentclass[10pt]{article}
\usepackage[letterpaper,margin=1in]{geometry}
\usepackage[T1]{fontenc}
\usepackage{microtype}
\usepackage{amsmath,amssymb,amsthm}
\usepackage{graphicx}
\usepackage{xcolor}
\usepackage{booktabs}
\usepackage{caption}
\usepackage[detect-all]{siunitx}
\usepackage{cite}
\usepackage{tikz}
\usetikzlibrary{arrows.meta,positioning}
\usepackage{pgfplots}
\pgfplotsset{compat=1.18}
\usepgfplotslibrary{fillbetween}
\pgfplotsset{every axis/.append style={tick label style={font=\footnotesize}, label style={font=\small},
  grid=major, grid style={black!12}, legend style={font=\footnotesize, draw=none, fill=none, cells={anchor=west}}}}
\usepackage[hidelinks]{hyperref}
\hypersetup{pdfauthor={Thammarat Yawisit (ORCID 0009-0006-9298-8862); Pittaya Pannil (ORCID 0000-0001-6082-3058)},
  pdftitle={Optimal Fan-Out and Break-Even Routing Loss in Waveguide-Integrated Ge-on-Si SPAD Arrays}}
\usepackage{orcidlink}

\newtheorem{proposition}{Proposition}
\newcommand{\Tobs}{\mathcal{T}}
\newcommand{\Bcal}{B$_\mathrm{cal}$}

\newif\ifieeenotice \ieeenoticetrue

\title{\bfseries Optimal Fan-Out and Break-Even Routing Loss\\ in Waveguide-Integrated Ge-on-Si SPAD Arrays}
\author{Thammarat Yawisit\,\orcidlink{0009-0006-9298-8862}\textsuperscript{*}
  \quad and \quad Pittaya Pannil\,\orcidlink{0000-0001-6082-3058}\\[4pt]
  \small Department of Instrumentation and Control Engineering, School of Engineering,\\
  \small King Mongkut's Institute of Technology Ladkrabang, Bangkok 10520, Thailand\\[2pt]
  \small \textsuperscript{*}Corresponding author: \href{mailto:tyawisit@icecube.wisc.edu}{tyawisit@icecube.wisc.edu};
  \ \href{mailto:pittaya.pa@kmitl.ac.th}{pittaya.pa@kmitl.ac.th}}
\date{}

\begin{document}
\maketitle
\ifieeenotice
\begingroup\renewcommand\thefootnote{}\footnotetext{\hspace{-1.8em}This work has been submitted to the IEEE for possible publication. Copyright may be transferred without notice, after which this version may no longer be accessible. This preprint is an extended version of a two-page conference paper.}\endgroup
\fi

\begin{abstract}
\noindent Splitting a waveguide mode over several Ge-on-Si single-photon avalanche diodes (SPADs) relieves dead-time saturation but multiplies dark counts. We quantify this trade-off with the asymptotic count-based information rate $\Gamma$, which prices dark counts and dead time in one unit and is attained by an optimally weighted moment estimator. For a non-paralyzable pixel, the information per pre-dead-time signal event peaks at a signal rate $y^\ast=\sqrt{D(D+1/\tau)}$ with exact efficiency $(\sqrt{1+D\tau}-\sqrt{D\tau})^2$, which sets the optimal fan-out $k^\ast\approx\eta T\Phi/y^\ast$ and a critical flux above which equal splitting is optimal. We then compare four routing architectures on 64 simulated 16-pixel arrays: an equal passive tree, static trees (uncalibrated, and an idealized calibration-informed benchmark), an ideal programmable tree, and a thermo-optic Mach--Zehnder interferometer (MZI) tree with \SI{1.32}{dB} excess loss. Ideal programmability gains up to \SI{15.9}{dB} over equal splitting at low flux. At one known flux, however, the calibrated static splitter beats the MZI tree by \SIrange{0.14}{2.7}{dB}, and the penalty tends to twice the excess loss when dark counts dominate. In ensemble-median worst-case regret, the MZI tree breaks even only for flux windows wider than 2.55 decades, and not within six decades above about \SI{3.4}{dB} excess loss. Event-level Monte Carlo simulations reproduce the count statistics and the predicted estimator variances within bootstrap confidence intervals. Sensitivity analyses cover dark-count level, heterogeneity, detector model and random seed. All results are simulations with representative literature parameters.
\end{abstract}


\section{Introduction}\label{sec:intro}
Germanium-on-silicon (Ge-on-Si) single-photon avalanche diodes (SPADs) extend photon counting on silicon photonic chips to the \SIrange{1.3}{1.55}{\micro\metre} telecommunication bands. Planar Ge-on-Si SPADs reach \SI{38}{\percent} single-photon detection efficiency at \SI{125}{K} and \SI{1310}{nm}~\cite{Vines2019}, GeSi SPADs have been operated at room temperature~\cite{Na2024}, and waveguide-coupled devices~\cite{Martinez2017} now reach \SI{34.6}{\percent} on-chip efficiency with \SI{279}{kHz} dark counts at \SI{78}{K}~\cite{Wang2023}. Two detector non-idealities limit what such a pixel can do with a bright optical mode: a dead time $\tau$ of order \SI{100}{ns} saturates the count rate near $1/\tau$, and dark-count rates $D$ of \SIrange{e4}{e6}{s^{-1}} bury weak signals.

The standard on-chip remedy for saturation is to split the mode over $N$ pixels. Actively switched $1{\times}N$ arrays outperform passive splitting trees in count rate~\cite{Castelletto2007}, but that analysis ignores dark counts and needs nanosecond switches. Splitting has a cost the count rate alone does not show: every added pixel adds its own dark counts, so a fan-out that is right for a bright input wastes information at low flux. Programmable photonic circuits built from thermo-optic Mach--Zehnder interferometers (MZIs)~\cite{Bogaerts2020,Harris2014} offer a middle way. They reconfigure in microseconds to milliseconds, so they set a power distribution rather than a per-photon route, and they add insertion loss. Separately, a reconfigurable circuit has been coupled directly to a visible-wavelength Si SPAD array~\cite{Gualandi2025}. These developments raise a design question this paper answers within an explicit model: \emph{when does slow, programmable redistribution of light over a Ge-on-Si SPAD array pay for its insertion loss?}

\paragraph{Contributions.} (i) Closed-form design quantities from the count-based information rate $\Gamma$: optimal load, exact efficiency, optimal fan-out, critical flux and dark-limited break-even loss (Section~\ref{sec:theory}). (ii) A comparison of four architectures on the same simulated arrays, with a calibration-informed static benchmark that separates calibration from reconfiguration (Section~\ref{sec:results}). (iii) Worst-case regret over flux windows and a map of the break-even window against excess loss (Section~\ref{sec:windows}). (iv) Event-level Monte Carlo validation and sensitivity analyses (Sections~\ref{sec:mc}--\ref{sec:sens}). This preprint extends a two-page conference paper with the same title and authors; its numerical results are those of the conference study, and every plotted series is read from the archived outputs.

\section{Detector, array and routing model}\label{sec:model}

\subsection{Pixel count statistics}
Pixel $i$ receives a Poisson stream of avalanche-triggering events at rate $x_i=y_i+D_i$, the sum of the pre-dead-time signal rate $y_i$ and the dark-count rate $D_i$. We model the dead time $\tau_i$ (including hold-off) as non-paralyzable: after each registered count, the pixel ignores all events for $\tau_i$. The registered counts then form a renewal process whose inter-count interval is $\tau_i$ plus an exponential variable of rate $x_i$, with mean $\tau_i+1/x_i$ and variance $1/x_i^2$. Renewal theory gives the asymptotic count rate and count variance per unit time~\cite{Muller1973,YuFessler2000},
\begin{equation}
R_i=\frac{x_i}{1+x_i\tau_i},\qquad \sigma_i^2=\lim_{\Tobs\to\infty}\frac{\mathrm{Var}[c_i]}{\Tobs}=\frac{\mathrm{Var}(\text{interval})}{\mathrm{E}(\text{interval})^3}=\frac{x_i}{(1+x_i\tau_i)^3},
\label{eq:rv}
\end{equation}
where $c_i$ is the number of counts in an observation time $\Tobs$. The Fano factor $\sigma_i^2/R_i=(1+x_i\tau_i)^{-2}$ falls below one: dead time regularizes the counts while it removes events. For a paralyzable pixel, in which every event restarts the dead time, $R=xe^{-x\tau}$ and $\sigma^2=xe^{-x\tau}(1-2x\tau e^{-x\tau})$; we use this variant only in the sensitivity analysis.

\subsection{Optical model}
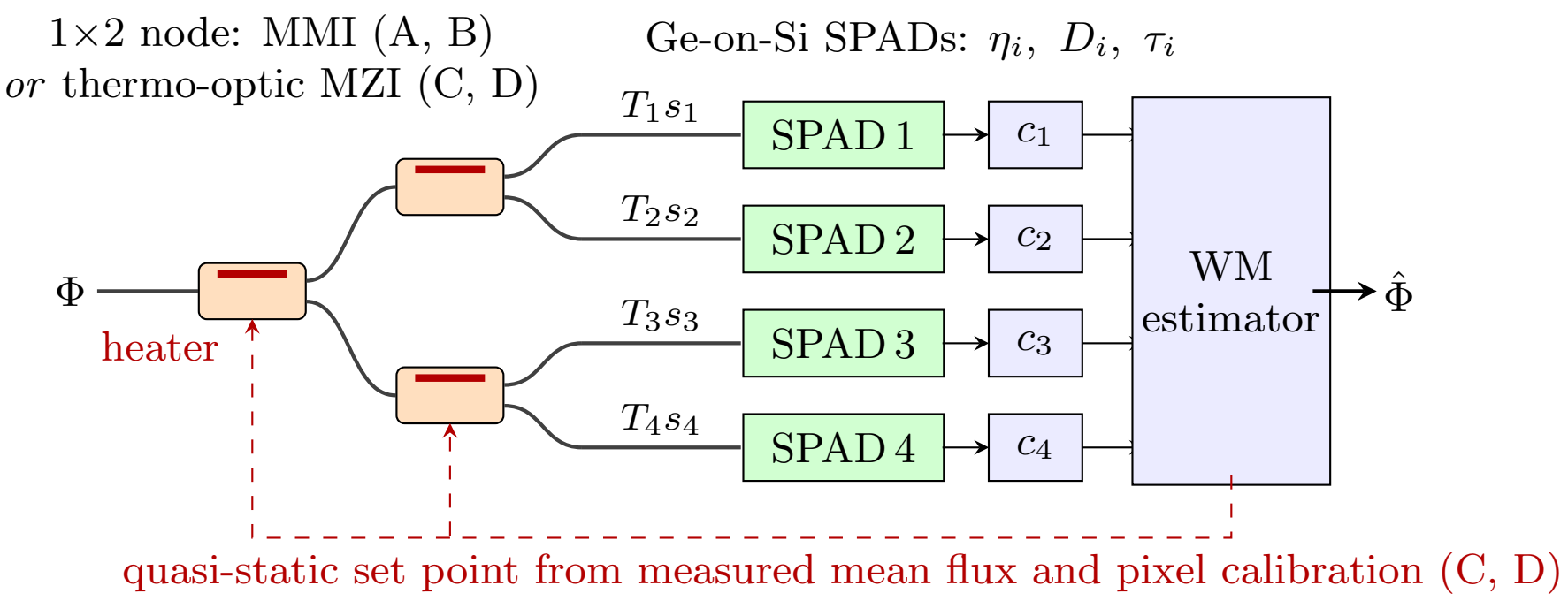
\begin{figure}[!htb]
\centering
\input{figs/fig_arch.tex}
\caption{Modeled architecture ($N=4$ drawn, $N=16$ analyzed). The input mode is distributed by a binary tree of $1{\times}2$ nodes, each a fixed multimode interference (MMI) splitter in architectures A and B or a thermo-optic MZI (heater in red) in C and D, over waveguide routes with transmission $T_i$ to Ge-on-Si SPADs with efficiency $\eta_i$, dark-count rate $D_i$ and dead time $\tau_i$. Counts $c_i$ feed a weighted-moment (WM) estimator of $\Phi$. For C and D, the split ratios are a quasi-static set point chosen from the measured mean flux and the calibrated pixel parameters, not a per-photon switch.}
\label{fig:arch}
\end{figure}

Light at photon rate $\Phi$ in the on-chip input waveguide (after the fiber coupler, whose loss is common to all architectures and omitted) reaches pixel $i$ with optical transmission $T_i$ and routing fraction $s_i$, $\sum_is_i=1$. With detection efficiency $\eta_i$, the pre-dead-time signal rate is
\begin{equation}
y_i=\eta_iT_is_i\Phi .
\label{eq:y}
\end{equation}
$T_i$ combines the insertion loss of the $m=\log_2N$ binary splitting stages with propagation loss along a route of length $L_i=\SI{2}{mm}+(i-1)\times\SI{0.05}{mm}$ at \SI{2}{dB/cm}. Routing fractions follow from a heap-ordered binary tree of split ratios $r_j$: each leaf fraction is the product of $r_j$ or $1-r_j$ along its path. Fig.~\ref{fig:arch} shows the topology for $N=4$; all results use $N=16$.

\subsection{Architectures}\label{sec:arch}
All architectures drive the same 16-pixel array, use the same routes and the same estimator (Section~\ref{sec:est}). They differ in how the $s_i$ are set, whether they use the measured pixel parameters, and in the per-stage loss.
\begin{itemize}\setlength\itemsep{1pt}
\item[(A)] \textbf{Equal passive tree.} A $1{\times}16$ MMI tree with an assumed \SI{0.1}{dB} excess loss per stage and fabrication split errors $r_j=0.5+\delta_j$, $\delta_j\sim\mathcal N(0,0.03^2)$, clipped to $[0.3,0.7]$.
\item[(B)] \textbf{Static trees.} B$_k$ is an uncalibrated $1{\times}k$ MMI tree ($\log_2k$ stages, the same split errors) on the first $k$ pixels, $k\in\{1,2,4,8,16\}$. B$^\ast$ picks, at each flux, the $k$ with the best ensemble-mean $\Gamma$: the designer knows the flux but not the individual pixels. \Bcal{} is an idealized calibration-informed static benchmark: one fixed splitting set, designed from measured pixel statistics, at assumed MMI-class loss. On-chip trimming would add MZI-like loss, so \Bcal{} is a bound on what calibration alone can achieve, not a demonstrated device. Its construction is given in Section~\ref{sec:bcal}.
\item[(C)] \textbf{Ideal programmable tree.} MMI-class loss, arbitrary $s$ on the simplex, re-optimized for each array and flux. C is an upper bound for any programmable network with the same per-stage loss.
\item[(D)] \textbf{Realistic MZI tree.} Each node is an MZI with insertion loss $\mathrm{IL}_\mathrm{MZI}=\SI{0.43}{dB}$: two MMIs at \SI{0.1}{dB} (assumed) plus a \SI{0.23}{dB} heater section~\cite{Harris2014}. The tree therefore has $4\times(0.43-0.10)=\SI{1.32}{dB}$ excess loss over A. A \SI{25}{dB} extinction ratio bounds each split ratio to $[\epsilon,1-\epsilon]$, $\epsilon=10^{-2.5}$. The 15 ratios are re-optimized for each array and flux.
\end{itemize}

\subsection{Simulated arrays}\label{sec:params}
Pixel parameters are drawn independently for each of 64 arrays (Table~\ref{tab:params}). These representative values come from devices of different structure and temperature~\cite{Vines2019,Na2024,Wang2023}; they do not describe one measured array. With $\sigma_{\ln D}=0.7$, one standard deviation changes the dark rate by a factor of two. All results are medians over the 64 arrays unless stated otherwise, with 16th--84th percentile bands.

\begin{table}[!htb]
\centering\small
\caption{Nominal simulation parameters (\texttt{code/config.json}).}
\label{tab:params}
\begin{tabular}{@{}lll@{}}
\toprule
Quantity & Value & Note \\
\midrule
Pixels per array, $N$ & 16 & 64 arrays, seed 20261023 \\
Detection efficiency $\eta_i$ & $0.20\,(1\pm0.10)$ & Gaussian, clipped to $[0.3,1.7]\times0.20$ \\
Dark-count rate $D_i$ & log-normal, median $D_0=\SI{e5}{s^{-1}}$, $\sigma_{\ln D}=0.7$ & variants $D_0=10^3$, $10^6$ \\
Dead time $\tau_i$ & $\SI{100}{ns}\,(1\pm0.05)$ & non-paralyzable \\
MMI excess loss & \SI{0.1}{dB} per stage & assumed \\
MZI insertion loss & \SI{0.43}{dB} per stage & $2\times0.1$ (assumed) $+\,0.23$~\cite{Harris2014} \\
Extinction ratio & \SI{25}{dB} & bounds the MZI ratios \\
Split error (MMI) & $\mathrm{sd}=0.03$ & clipped to $[0.3,0.7]$ \\
Routes & \SIrange{2}{2.75}{mm} at \SI{2}{dB/cm} & mean transmission $T=0.818$ incl.\ MMIs \\
Flux grid & $10^4$--\SI{e10}{s^{-1}}, 37 points & 1/6-decade spacing \\
\bottomrule
\end{tabular}
\end{table}

\section{Information-rate theory}\label{sec:theory}

\subsection{Figure of merit and estimator}\label{sec:est}
For independent pixels and large $\Tobs$, the counts are jointly Gaussian with means $\Tobs R_i(\Phi)$ and variances $\Tobs\sigma_i^2$. With $y_i=a_i\Phi$, $a_i=\eta_iT_is_i$, we have $\partial R_i/\partial\Phi=a_i/(1+x_i\tau_i)^2$, and the Fisher information on $\ln\Phi$ per unit time is
\begin{equation}
\Gamma\equiv\Phi^2\sum_i\frac{(\partial R_i/\partial\Phi)^2}{\sigma_i^2}=\sum_i g_i(y_i),\qquad g_i(y)=\frac{y^2}{(y+D_i)\,[1+(y+D_i)\tau_i]} .
\label{eq:gamma}
\end{equation}
$\Gamma$ has units of s$^{-1}$: it is the count rate of an ideal, dark-free, dead-time-free counter that would give the same relative precision. For such a counter $\Gamma=y$. The weighted-moment estimator $\hat\Phi$ that solves
\begin{equation}
\sum_iw_i\Bigl[\frac{c_i}{\Tobs}-R_i(\hat\Phi)\Bigr]=0,\qquad w_i=\frac{\partial R_i/\partial\Phi}{\sigma_i^2},
\label{eq:wm}
\end{equation}
is the optimally weighted (generalized-method-of-moments) estimator for the count vector and reaches $\mathrm{sd}(\hat\Phi)/\Phi=(\Gamma\Tobs)^{-1/2}$ asymptotically~\cite{Kay1993}. We do not claim it is the maximum-likelihood estimator. Because any reweighting or exclusion of count channels ($w_i=0$) is a special case of \eqref{eq:wm}, none can exceed $\Gamma$. Arrival-time information is outside this count-only model: $\Gamma$ is the information carried by the counts, not by time stamps. We therefore call $\Gamma$ the asymptotic count-based information rate, and $\Gamma/\Phi$ the information efficiency.

\subsection{Optimal load and exact efficiency}
The ratio $g(y)/y$ is the information per pre-dead-time signal event. It is small at low load, where dark counts dominate ($g\approx y^2/[D(1+D\tau)]$, convex), and at high load, where dead time saturates the pixel ($g\to1/\tau$, concave).

\begin{figure}[!htb]
\centering
\input{figs/fig_analytic.tex}
\caption{Optimal pixel load (closed form, $\tau=\SI{100}{ns}$). (a) Information per pre-dead-time signal event $g(y)/y$ for $D=10^3$--\SI{e6}{s^{-1}}; dots mark $y^\ast$ from \eqref{eq:yopt}. (b) Normalized optimal load $y^\ast\tau$ and peak efficiency versus $D\tau$ (lines), with the numerical optima for a paralyzable pixel (markers, E5).}
\label{fig:analytic}
\end{figure}
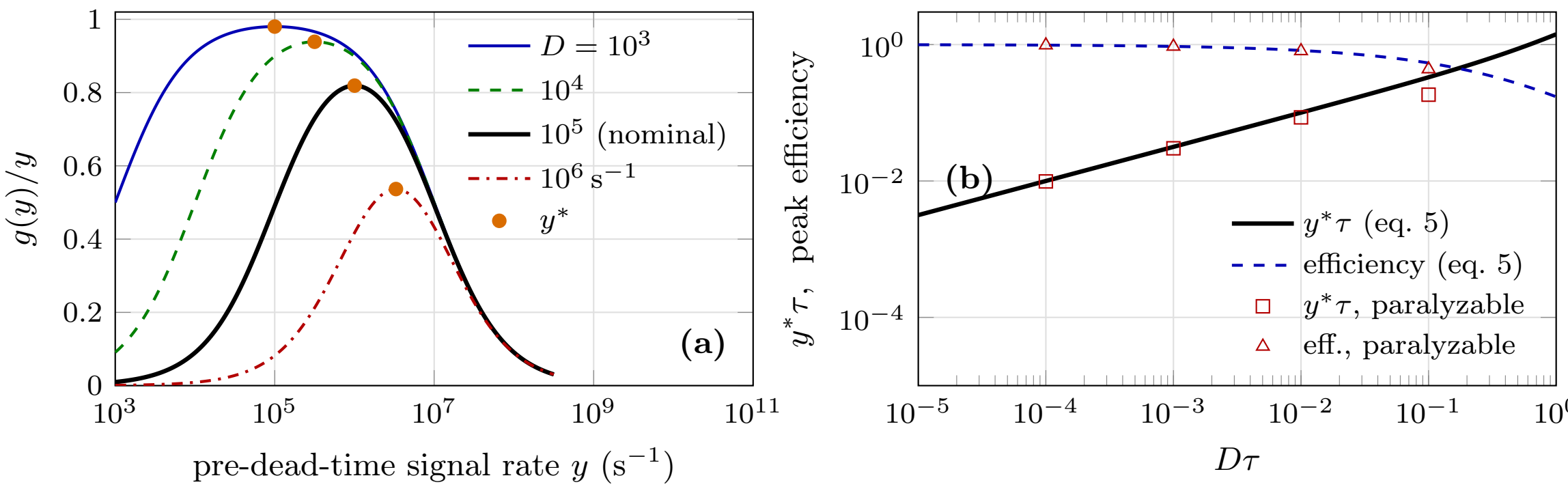

\begin{proposition}[Optimal load]\label{prop:yopt}
For a non-paralyzable pixel with dark-count rate $D$ and dead time $\tau$, $g(y)/y$ has a unique maximum on $y>0$ at
\begin{equation}
y^\ast=\sqrt{D\,(D+1/\tau)},\qquad \frac{g(y^\ast)}{y^\ast}=\Bigl(\sqrt{1+D\tau}-\sqrt{D\tau}\Bigr)^2 .
\label{eq:yopt}
\end{equation}
\end{proposition}
\begin{proof}
$\ln[g(y)/y]=\ln y-\ln(y+D)-\ln[1+(y+D)\tau]$. Its derivative $1/y-1/(y+D)-\tau/[1+(y+D)\tau]$ vanishes when $(y+D)[1+(y+D)\tau]=y[1+2(y+D)\tau]$, that is $\tau y^2=D+D^2\tau$, which has one positive root. The derivative is positive below and negative above it. With $u=D\tau$, substitution gives $g(y^\ast)/y^\ast=(\sqrt u+\sqrt{1+u})^{-2}$, and $(\sqrt{1+u}-\sqrt u)(\sqrt{1+u}+\sqrt u)=1$ gives \eqref{eq:yopt}.
\end{proof}
For the nominal pixel ($D=\SI{e5}{s^{-1}}$, $\tau=\SI{100}{ns}$, $D\tau=0.01$), $y^\ast=\SI{1.005e6}{s^{-1}}$ and the peak efficiency is 0.819. A dense grid maximum agrees with $y^\ast$ to $2\times10^{-6}$, and SymPy confirms \eqref{eq:gamma} and \eqref{eq:yopt} symbolically (\texttt{results/derivation\_checks.json}). For $D\tau\ll1$, $y^\ast\approx\sqrt{D/\tau}$, the geometric mean of the dark rate and the saturation rate, and the efficiency is $\approx1-2\sqrt{D\tau}$. Fig.~\ref{fig:analytic}(a) shows $g(y)/y$ for four dark rates, and Fig.~\ref{fig:analytic}(b) the dependence of $y^\ast\tau$ and the peak efficiency on $D\tau$: at $D=\SI{e6}{s^{-1}}$ the best a pixel can do is 0.54 of an ideal counter.

\subsection{Optimal fan-out and critical flux}\label{sec:fanout}
For $N$ identical pixels with transmission $T$ and total pre-dead-time signal rate $Y=\eta T\Phi$, the allocation problem $\max\sum_ig(y_i)$ subject to $\sum_iy_i=Y$ has a convex--concave objective. Its solution loads $k$ pixels equally and leaves the rest dark, with
\begin{equation}
k^\ast\approx\frac{\eta T\Phi}{y^\ast}\ \text{clipped to}\ [1,N],\qquad \Phi_c=\frac{N\,y^\ast}{\eta T},
\label{eq:kstar}
\end{equation}
so equal splitting over all $N$ pixels is optimal only above the critical flux $\Phi_c$. With the nominal mean transmission $T=0.818$, $\Phi_c=\SI{9.8e7}{s^{-1}}$. Below $\Phi_c$, every pixel beyond $k^\ast$ adds dark counts faster than it relieves dead time. Heterogeneous pixels change the picture in one respect: the optimizer prefers pixels with a high $\eta_i^2T_i^2/[D_i(1+D_i\tau_i)]$, so dark-count spread creates a selection gain on top of the fan-out gain (Section~\ref{sec:fanres}).

\subsection{Dark-limited break-even loss}\label{sec:darkbe}
When $y\ll D$ for every active pixel, $g\approx y^2/[D(1+D\tau)]\propto T^2$. An architecture that gains $G$\,dB in $\Gamma$ over another therefore loses that gain to an extra insertion loss of
\begin{equation}
\Delta\mathrm{IL}_\mathrm{be}=G/2\ \text{dB} .
\label{eq:be}
\end{equation}
For a uniform array, concentrating all light on one pixel instead of 16 gains $10\log_{10}16=\SI{12.04}{dB}$, so the break-even excess loss is \SI{6.02}{dB}. The same scaling explains why loss costs a programmable tree twice its value in dB at low flux (Section~\ref{sec:static}). When pixels are saturated, $g\to(1/\tau)\,y^2/x^2$ depends only weakly on $T$, and loss costs little information.

\subsection{Afterpulsing and gated operation}
Afterpulses are correlated with preceding detections, so \eqref{eq:rv} does not describe them. If they are crudely treated as an extra uncorrelated dark rate $D_\mathrm{eff}>D$, then $\partial y^\ast/\partial D=(2D+1/\tau)/(2y^\ast)>0$ and the optimal load \emph{rises}. A quantitative treatment needs a temporal model and is outside this study. For gated operation with gate width $t_g$, the free-running results hold with $\tau\to t_g/2$ and $D$ unchanged, to within 1--\SI{7}{\percent} for a dark mean per gate $Dt_g\le0.02$ (Appendix~\ref{app:variants}).

\section{Computational pipeline}\label{sec:pipeline}

\begin{figure}[!htb]
\centering
\input{figs/fig_pipeline.tex}
\caption{Computational pipeline. Grey: inputs and model; orange: analysis and optimization; blue: evaluation; green: event-level validation; red: outputs. Every figure and table in this paper reads a file produced by this chain (\texttt{code/run\_all.sh}, \texttt{code/make\_arxiv\_figdata.py}).}
\label{fig:pipeline}
\end{figure}
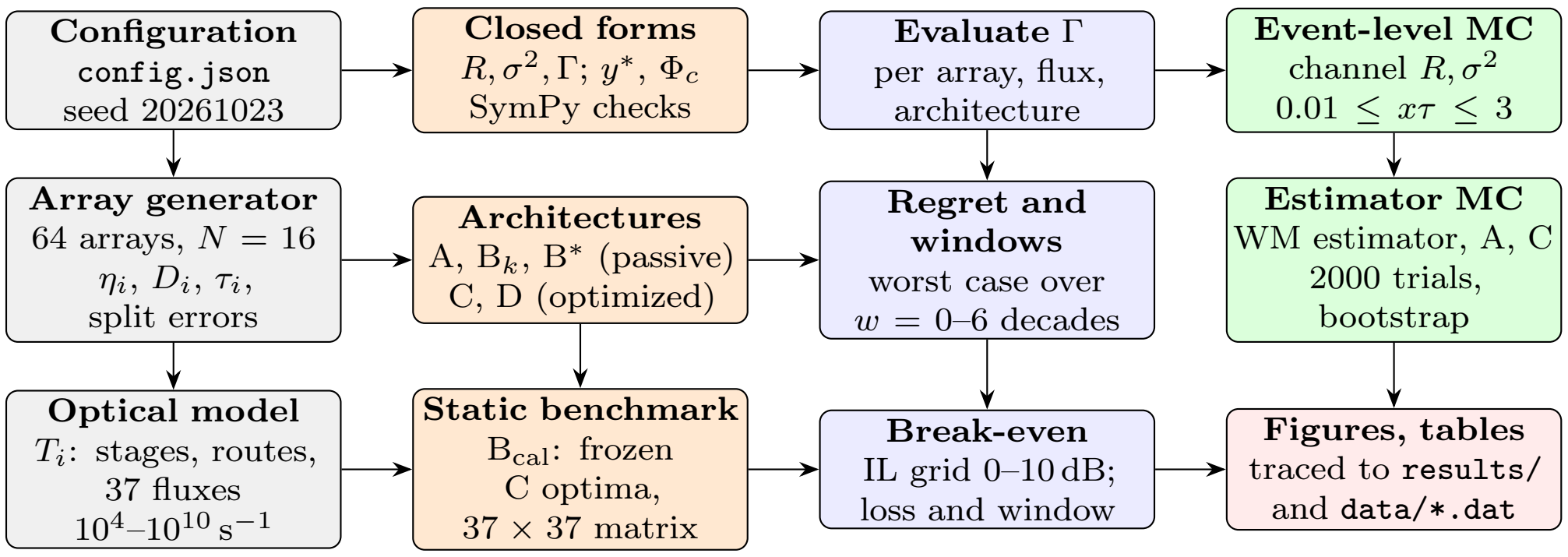

Fig.~\ref{fig:pipeline} summarizes the computation, from the configuration of Table~\ref{tab:params} to the figures.

\paragraph{Optimizers.} C solves $\max_s\sum_ig_i(a_is_i\Phi)$ on the simplex. The problem is not concave, so we use 16 structured starts (equal split over the $k$ pixels with the highest $g_i$ at load $1/k$, $k=1,\dots,16$) and polish the best three with sequential least-squares programming (SLSQP). D maps the optimum of C onto the tree, clips the ratios to the extinction bounds, and polishes with L-BFGS-B from that point and from the equal split. A check against 60 random restarts on 16 arrays and five fluxes found no shortfall, and exhaustive search on a simplex grid for $N=4$ found a largest shortfall of \SI{0.6}{\percent} (\SI{0.026}{dB}, E6).

\paragraph{Calibrated static benchmark.}\label{sec:bcal} For each array, the candidate static designs are the optima of C at the 37 grid fluxes, each frozen and evaluated at every flux. This gives a $37\times37$ matrix of $\Gamma$ per array (\texttt{E7\_static\_matrix.npz}). For a single known flux, \Bcal{} is the candidate designed at that flux and equals C exactly (largest deviation on the diagonal: 0.0). For a flux window, \Bcal{} is the candidate with the smallest worst-case regret over the window, chosen per array. Because \Bcal{} shares the splitting set and loss model of C, the comparison D versus \Bcal{} isolates the value of reconfiguration from that of calibration.

\paragraph{Regret and flux windows.} The regret of a design at flux $\Phi$ is $\rho(\Phi)=10\log_{10}[\Gamma_\mathrm{C}(\Phi)/\Gamma(\Phi)]\ge0$. For a window $W=[\Phi_0,\Phi_010^w]$ on the 1/6-decade grid, the worst-case regret is $\max_{\Phi\in W}\rho$. B$_k$ chooses the $k$ that minimizes the median over arrays of the worst-case regret, since its designer does not know the array. Statistics are taken over arrays and window positions (2368 cases at $w=0$).

\paragraph{Break-even loss and window.} For the break-even loss, $\Gamma_\mathrm{D}$ is evaluated on 26 per-stage insertion losses (0.1--\SI{2.6}{dB}, i.e.\ 0--\SI{10}{dB} excess), and the crossing with the reference is interpolated linearly in $\log\Gamma$. For the break-even window, we find the smallest $w$ at which the median worst-case regret of D falls below that of the static baseline, interpolated linearly in $w$. The event-level Monte Carlo (MC) generates Poisson time stamps and registers counts event by event under the dead-time rule, independently of \eqref{eq:rv}. The full run takes about one hour on four cores.

\section{Results}\label{sec:results}

\subsection{Single-flux performance}
\begin{figure}[!htb]
\centering
\input{figs/fig_performance.tex}
\caption{Single-flux performance, medians over 64 simulated arrays. (a) Information efficiency $\Gamma/\Phi$; band: 16th--84th percentile of D; \Bcal: one static calibrated design per array, minimax over the plotted range; markers: event-level MC on one array (Section~\ref{sec:mc}). (b) Gain over A for C and D (bands: 16th--84th percentile) and for the flux-aware uncalibrated fan-out B$^\ast$.}
\label{fig:perf}
\end{figure}
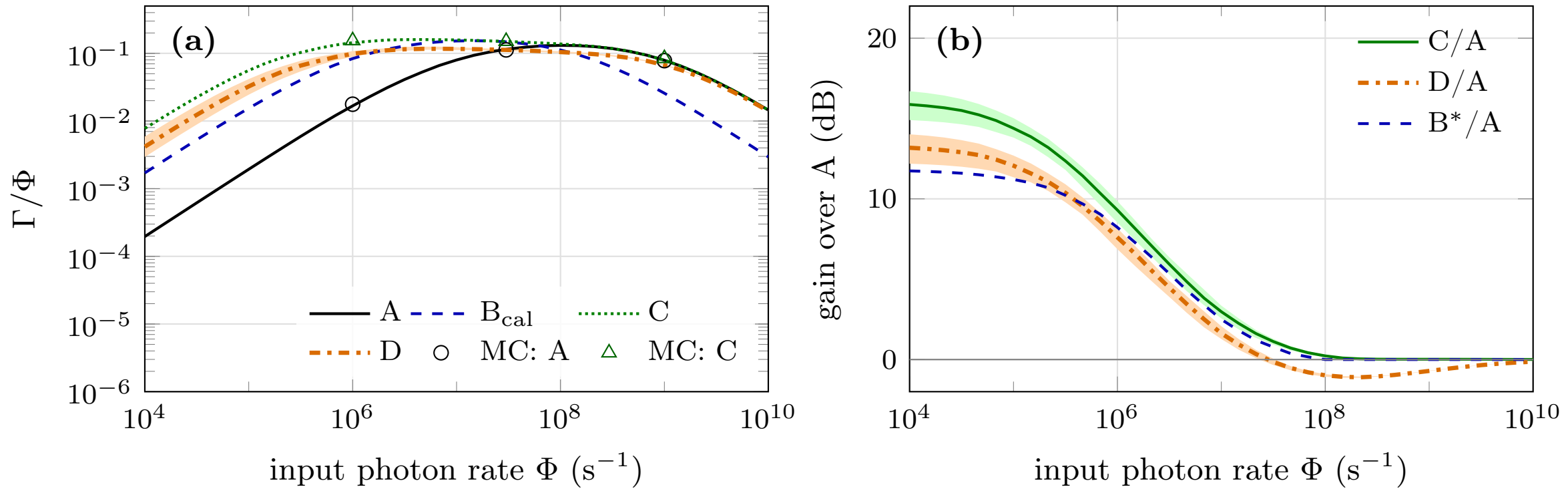

Fig.~\ref{fig:perf}(a) shows the information efficiency $\Gamma/\Phi$ of each architecture. The equal tree A is dark-count limited below about \SI{e7}{s^{-1}}: each of its 16 pixels receives 1/16 of the signal but its full dark rate. The ideal programmable tree C concentrates light on the best pixels and gains up to \SI{15.9}{dB} over A at \SI{e4}{s^{-1}} [Fig.~\ref{fig:perf}(b)]. The gain falls to \SI{3.0}{dB} at \SI{e7}{s^{-1}} and \SI{0.02}{dB} at \SI{e9}{s^{-1}}, where equal splitting is optimal. The MZI tree D follows C with a loss penalty: it gains up to \SI{13.2}{dB} over A at low flux, but falls below A above $\approx\SI{2.8e7}{s^{-1}}$, by up to \SI{1.1}{dB} near \SI{2e8}{s^{-1}}. The flux-aware uncalibrated tree B$^\ast$ captures most of the fan-out gain (\SI{11.7}{dB} at \SI{e4}{s^{-1}}) but none of the selection gain. Table~\ref{tab:rep} lists representative values.

\begin{table}[!htb]
\centering\small
\caption{Representative fluxes (medians over 64 arrays; \texttt{results/table\_representative\_fluxes.tex}). $n_\mathrm{C}$: pixels used by C; $k_{\mathrm{B}^\ast}$: fan-out of B$^\ast$; $G$: gain over A; $\Delta\mathrm{IL}_\mathrm{be}$: excess loss of D at which it falls to A or B$^\ast$.}
\label{tab:rep}
\begin{tabular}{@{}rrrrrrrr@{}}
\toprule
$\Phi$ (s$^{-1}$) & $n_\mathrm{C}$ & $k_{\mathrm{B}^\ast}$ & $G_\mathrm{C}$ (dB) & $G_\mathrm{D}$ (dB) & D vs B$^\ast$ (dB) & $\Delta\mathrm{IL}_\mathrm{be}$ vs A (dB) & vs B$^\ast$ (dB) \\
\midrule
$10^{5}$ & 1 & 1 & 14.4 & 12.1 & $+$0.8 & 7.86 & 1.82 \\
$10^{7}$ & 2 & 2 & 3.0 & 1.6 & $-$1.0 & 3.00 & 0.34 \\
$10^{9}$ & 16 & 16 & 0.0 & $-$0.7 & $-$0.7 & 0.03 & 0.03 \\
\bottomrule
\end{tabular}
\end{table}

\subsection{Fan-out, pixel selection and heterogeneity}\label{sec:fanres}
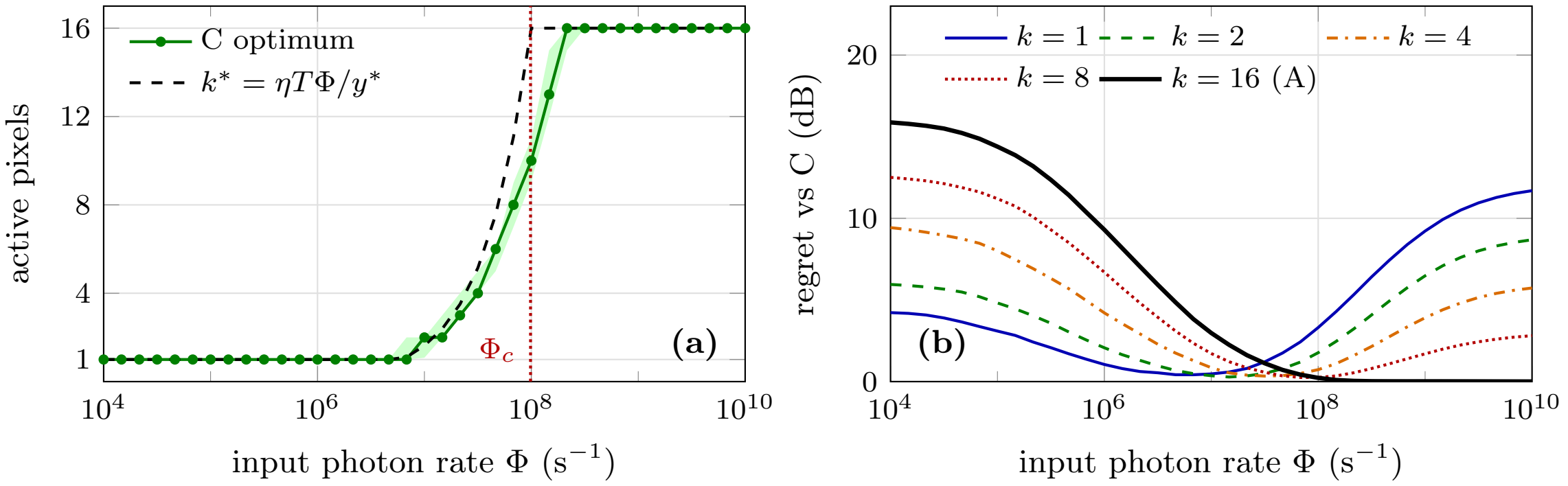
\begin{figure}[!htb]
\centering
\input{figs/fig_fanout.tex}
\caption{(a) Number of pixels used by the optimum of C ($s_i>10^{-3}$; median and 16th--84th percentile band) and the identical-pixel estimate $k^\ast$ of \eqref{eq:kstar}; dotted: $\Phi_c$. (b) Median regret relative to C of uncalibrated fixed $1{\times}k$ trees; $k=16$ is architecture A.}
\label{fig:fanout}
\end{figure}

The optimum of C uses a single pixel below \SI{e7}{s^{-1}} and all 16 above \SI{2e8}{s^{-1}} [Fig.~\ref{fig:fanout}(a)]. The transition tracks the identical-pixel estimate $k^\ast$ of \eqref{eq:kstar} and completes near $\Phi_c$. Fig.~\ref{fig:fanout}(b) shows why no single passive fan-out serves the whole range. A one-pixel tree loses \SI{4.2}{dB} to C at \SI{e4}{s^{-1}}, because its pixel is not the best one, comes within \SI{0.5}{dB} near \SI{e7}{s^{-1}}, and loses \SI{11.7}{dB} at \SI{e10}{s^{-1}}; the 16-pixel tree loses \SI{15.9}{dB} at low flux and almost nothing at high flux. Relative to the flux-aware B$^\ast$, the best single fixed fan-out ($k=4$) still loses up to \SI{5.7}{dB} somewhere in the range.

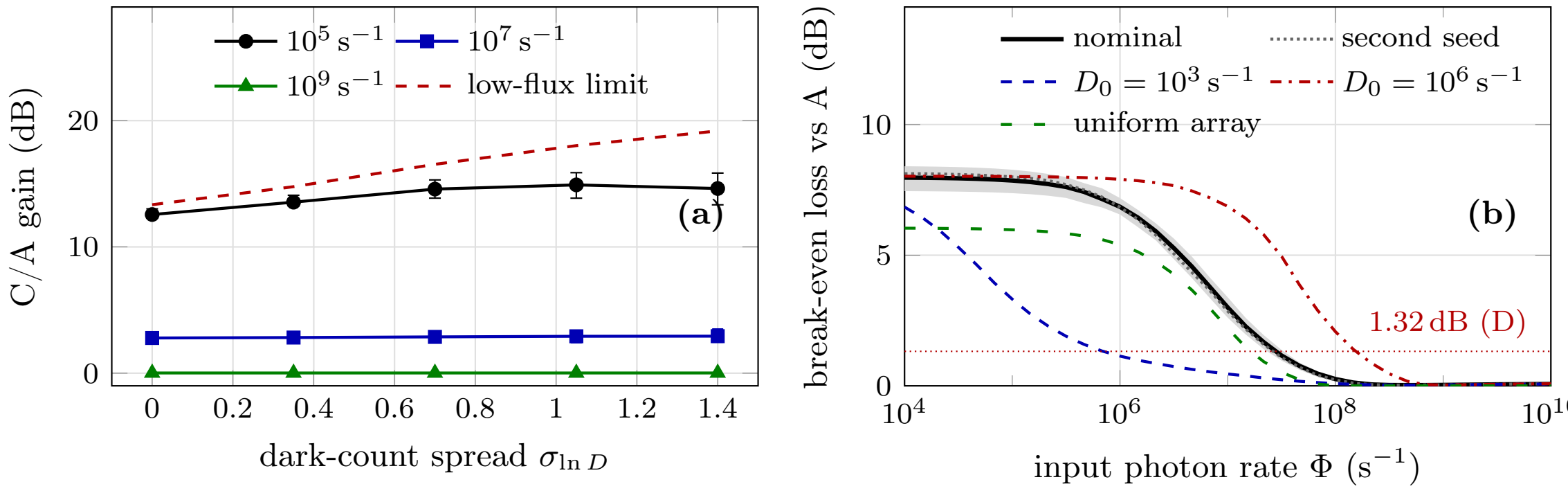
\begin{figure}[!htb]
\centering
\input{figs/fig_robust.tex}
\caption{(a) C/A gain versus dark-count spread at three fluxes (E3; median, bars: 16th--84th percentile; the \SI{e9}{s^{-1}} bars are smaller than the markers) and the low-flux closed form (dashed). (b) Excess loss at which D falls to A (median; band: 16th--84th percentile, nominal) for the nominal ensemble and four variants; dotted: the \SI{1.32}{dB} of the modeled MZI tree.}
\label{fig:robust}
\end{figure}

The gain of C over A has two parts. For a uniform array, it is pure fan-out: \SI{12.1}{dB} at low flux, close to $10\log_{10}16$. The remainder, about \SI{3.8}{dB} at the low-flux maximum of the nominal ensemble, comes from selecting the best pixels. Fig.~\ref{fig:robust}(a) varies the dark-count spread at fixed median (E3, 64 arrays per point). At \SI{e5}{s^{-1}}, the C/A gain rises from \SI{12.6}{dB} for identical dark rates to \SI{14.9}{dB} at $\sigma_{\ln D}=1.05$, then saturates. The low-flux closed form $10\log_{10}[N\max_iq_i/\overline{q}]$ with $q_i=\eta_i^2/[D_i(1+D_i\tau_i)]$ keeps rising, because at \SI{e5}{s^{-1}} the best pixel is no longer far below its optimal load. At \SI{e7}{s^{-1}} heterogeneity adds only about \SI{0.1}{dB}, and at \SI{e9}{s^{-1}} nothing.

\subsection{Break-even loss against the equal tree}
Fig.~\ref{fig:robust}(b) shows how much excess loss D can carry before it falls to A. At low flux the nominal break-even is \SI{7.97}{dB}, half of the \SI{15.9}{dB} C/A gain, as \eqref{eq:be} predicts; for the uniform array it is \SI{6.03}{dB}, against \SI{6.02}{dB} predicted. The break-even drops through the modeled \SI{1.32}{dB} near \SI{2.8e7}{s^{-1}} and is 0.03--\SI{0.07}{dB} above \SI{3e8}{s^{-1}}. A second random seed (32 arrays) moves the curve by at most \SI{0.22}{dB}. Dark rate shifts the curve along the flux axis: with $D_0=\SI{e3}{s^{-1}}$ the crossing of \SI{1.32}{dB} moves down to $\approx\SI{7e5}{s^{-1}}$, and with $D_0=\SI{e6}{s^{-1}}$ (32 arrays) up to $\approx\SI{1.5e8}{s^{-1}}$.

\subsection{Calibrated static benchmark at one flux}\label{sec:static}
At a single known flux, \Bcal{} attains the optimum of C, so D can only lose. Fig.~\ref{fig:windows}(a) shows by how much: the median penalty runs from \SI{2.69}{dB} at \SI{e4}{s^{-1}} to \SI{0.14}{dB} at \SI{e10}{s^{-1}}, and no simulated array at any flux favors D (best case \SI{-0.13}{dB}). The low-flux value is close to twice the \SI{1.32}{dB} excess loss, the dark-limited scaling of Section~\ref{sec:darkbe}; the small remainder is consistent with the extinction-ratio bound, which leaks some light to unused pixels. At high flux the saturated pixels are insensitive to loss. The \SI{1.5}{dB} advantage of D over B$^\ast$ at \SI{e4}{s^{-1}} (Table~\ref{tab:rep}) is therefore the value of calibration, not of programmability.

\subsection{Flux windows and break-even width}\label{sec:windows}
\begin{figure}[!htb]
\centering
\input{figs/fig_windows.tex}
\caption{(a) D relative to \Bcal{} designed at the evaluation flux (equal to C there): median, 16th--84th percentile band and best of 64 arrays; dotted: one and two times the \SI{1.32}{dB} excess loss. (b) Ensemble-median worst-case regret versus flux-window width for \Bcal, B$_k$ and D at three excess losses; bars: 16th--84th percentile over arrays and window positions at nine widths (left \Bcal, right D at \SI{1.32}{dB}). (c) Smallest window width for which D has the lower median worst-case regret, versus its excess loss; no point: no crossover within six decades.}
\label{fig:windows}
\end{figure}
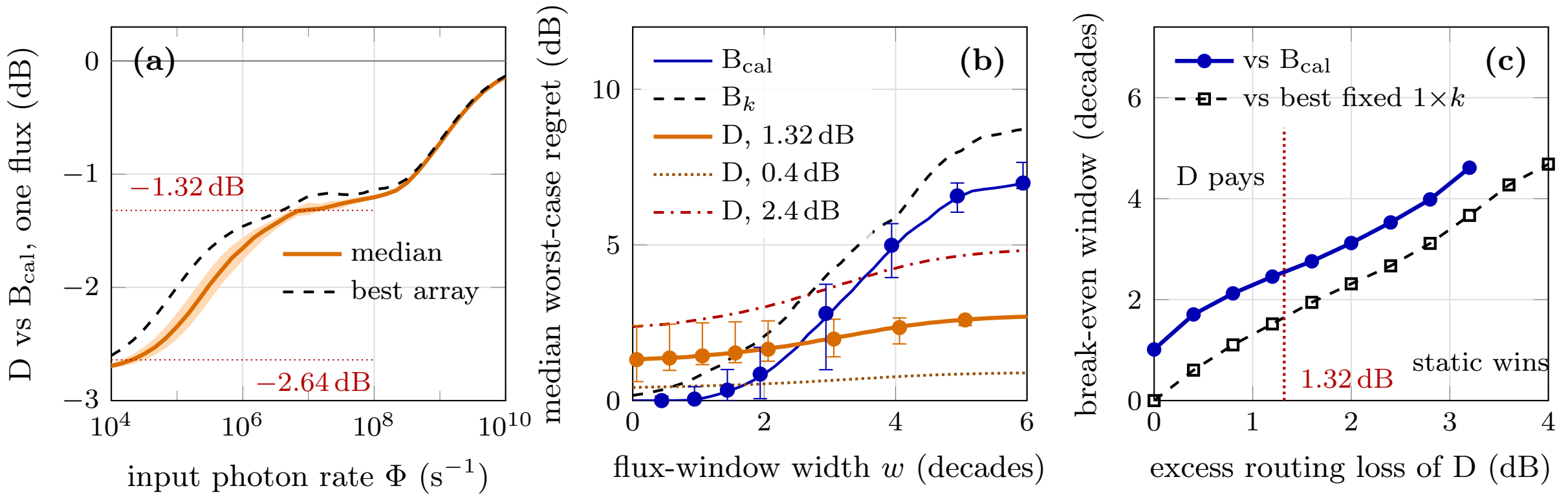

Reconfiguration can only pay when one fixed design must serve a range of fluxes. Fig.~\ref{fig:windows}(b) and Table~\ref{tab:regret} give the median worst-case regret against the window width. \Bcal{} has no regret at $w=0$, but its regret grows quickly with width: \SI{0.85}{dB} at two decades, \SI{2.80}{dB} at three and \SI{6.99}{dB} at six. The uncalibrated B$_k$ is worse at every width. D starts at about its excess loss and grows slowly, to \SI{2.69}{dB} at six decades for \SI{1.32}{dB} excess. Individual arrays scatter: at two decades, the 16th--84th percentile ranges are 0.06--\SI{1.71}{dB} for \Bcal{} and 1.27--\SI{2.56}{dB} for D.

\begin{table}[!htb]
\centering\small
\caption{Median worst-case regret (dB) over arrays and window positions (\texttt{results/E7\_summary.json}).}
\label{tab:regret}
\begin{tabular}{@{}lrrrrrrrrr@{}}
\toprule
Window width (decades) & 0 & 0.5 & 1 & 1.5 & 2 & 3 & 4 & 5 & 6 \\
\midrule
\Bcal & 0.00 & 0.00 & 0.05 & 0.34 & 0.85 & 2.80 & 4.99 & 6.57 & 6.99 \\
B$_k$ & 0.17 & 0.37 & 0.79 & 1.37 & 2.07 & 4.13 & 5.86 & 8.03 & 8.73 \\
D, \SI{0.4}{dB} excess & 0.42 & 0.44 & 0.47 & 0.50 & 0.54 & 0.64 & 0.77 & 0.85 & 0.89 \\
D, \SI{1.32}{dB} excess & 1.32 & 1.37 & 1.44 & 1.53 & 1.65 & 1.98 & 2.35 & 2.59 & 2.69 \\
D, \SI{2.4}{dB} excess & 2.38 & 2.45 & 2.60 & 2.77 & 3.00 & 3.61 & 4.25 & 4.65 & 4.82 \\
\bottomrule
\end{tabular}
\end{table}

Fig.~\ref{fig:windows}(c) maps the crossover. At the modeled \SI{1.32}{dB}, the ensemble median favors D beyond \num{2.55} decades against \Bcal{} and \num{1.63} decades against B$_k$. Even lossless reconfiguration needs \num{1.01} decades against \Bcal, because a calibrated static design is nearly optimal over about one decade. The required width grows with loss: 1.71 decades at \SI{0.4}{dB}, 3.12 at \SI{2.0}{dB} and 4.61 at \SI{3.2}{dB}. On the studied \SI{0.4}{dB} grid, D has no break-even within six decades beyond 3.2--\SI{3.6}{dB} excess loss against \Bcal, and beyond 4.0--\SI{4.4}{dB} against B$_k$.

\section{Monte Carlo validation}\label{sec:mc}
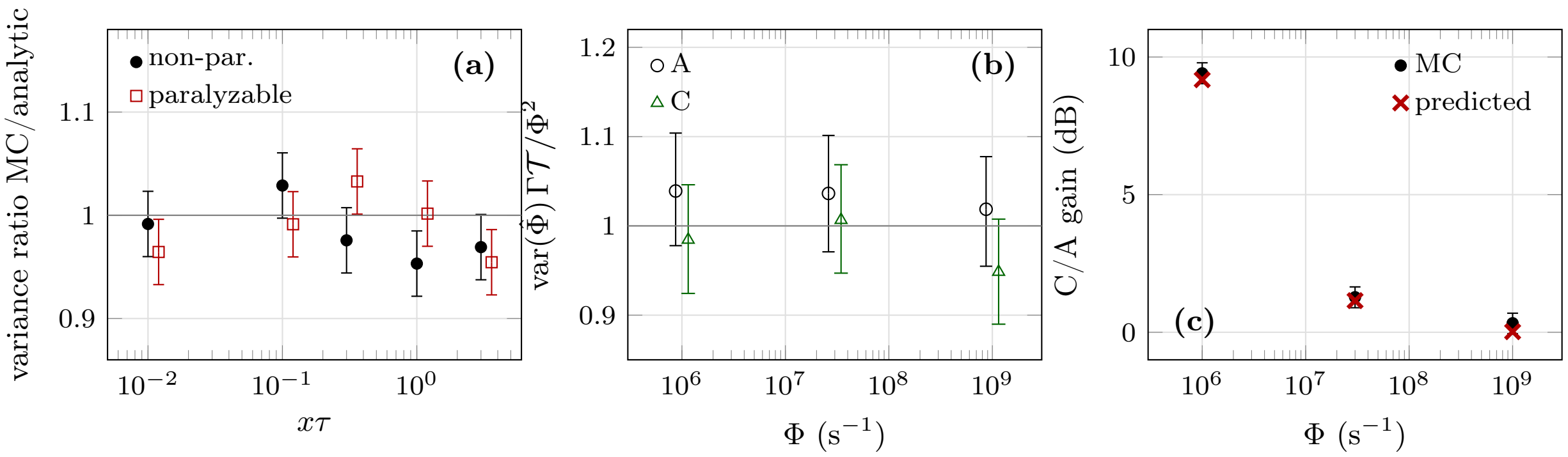
\begin{figure}[!htb]
\centering
\input{figs/fig_mc.tex}
\caption{Event-level Monte Carlo validation. (a) Simulated over analytic count variance for one channel (E4a; bars: $\pm1$ standard error). (b) Estimator variance relative to $1/(\Gamma\Tobs)$ for A (circles) and C (triangles) on one simulated array (E4d, 2000 trials; bars: bootstrap 95\% confidence intervals; markers offset horizontally for clarity). (c) MC-estimated C/A gain with bootstrap 95\% intervals and the analytic prediction.}
\label{fig:mc}
\end{figure}

\paragraph{Channel statistics (E4a).} For $x\tau\in\{0.01,0.1,0.3,1,3\}$, 2000 trials of about 3000 registered counts each reproduce the analytic mean count of \eqref{eq:rv} within \SI{0.1}{\percent} and the variance within 1.5 standard errors (ratios 0.95--1.03, standard error 0.032), for both the non-paralyzable and the paralyzable rule [Fig.~\ref{fig:mc}(a)].

\paragraph{Estimator variance (E4d).} On the first simulated array of the nominal ensemble, at $\Phi=10^6$, $3\times10^7$ and \SI{e9}{s^{-1}}, we simulate all 16 channels of A and of C (at its optimum) for 2000 trials and solve \eqref{eq:wm} for each trial. The observation time gives about 3000 counts on the busiest pixel. Table~\ref{tab:mc} and Fig.~\ref{fig:mc}(b) show that the estimator variance relative to $\Phi^2/(\Gamma\Tobs)$ is 0.95--1.04, and all six bootstrap 95\% confidence intervals ($B=4000$) contain 1. The relative bias is below \SI{0.4}{\percent} in every case and within two standard errors of zero. The MC-estimated C/A gains of 9.41, 1.28 and \SI{0.33}{dB} have 95\% intervals of about $\pm\SI{0.38}{dB}$ that contain the predictions of 9.17, 1.15 and \SI{0.02}{dB} [Fig.~\ref{fig:mc}(c)]. One bootstrap standard error is about \SI{0.19}{dB} per gain, so the residual differences (at most \SI{0.31}{dB}) are within two standard errors.

\begin{table}[!htb]
\centering\small
\caption{Estimator Monte Carlo on array 1: 2000 trials per case, bootstrap $B=4000$ (\texttt{E4d\_mc\_bootstrap.json}).}
\label{tab:mc}
\begin{tabular}{@{}rccccc@{}}
\toprule
$\Phi$ (s$^{-1}$) & $n_\mathrm{C}$ & var/CRB, A [95\% CI] & var/CRB, C [95\% CI] & C/A gain, MC [95\% CI] & predicted \\
\midrule
$10^6$ & 1 & 1.039 [0.978, 1.104] & 0.984 [0.924, 1.046] & \SI{9.41}{dB} [9.03, 9.79] & \SI{9.17}{dB} \\
$3\times10^7$ & 4 & 1.036 [0.971, 1.101] & 1.007 [0.947, 1.069] & \SI{1.28}{dB} [0.89, 1.65] & \SI{1.15}{dB} \\
$10^9$ & 16 & 1.019 [0.955, 1.078] & 0.949 [0.890, 1.008] & \SI{0.33}{dB} [$-0.05$, 0.69] & \SI{0.02}{dB} \\
\bottomrule
\end{tabular}
\end{table}

\section{Sensitivity and robustness}\label{sec:sens}
Table~\ref{tab:sens} collects the headline quantities for the nominal ensemble and four variants. The cooled variant ($D_0=\SI{e3}{s^{-1}}$) reduces the C/A gain to \SI{9.4}{dB}, because the dark-limited regime shrinks to fluxes below \SI{e5}{s^{-1}}, and D never beats the flux-aware B$^\ast$ (best \SI{-0.13}{dB}). The hot variant ($D_0=\SI{e6}{s^{-1}}$) extends the dark-limited regime and lets D beat B$^\ast$ by up to \SI{2.0}{dB}. For a uniform array, the selection gain disappears and D never beats B$^\ast$ (best \SI{-0.14}{dB}): with identical pixels, the right fan-out is all that routing can provide, and a passive tree provides it with less loss. The second seed changes the maxima by at most \SI{0.3}{dB}. The paralyzable model lowers $y^\ast$ by \SI{15}{\percent} at $D\tau=0.01$ (Fig.~\ref{fig:analytic}b, Appendix~\ref{app:variants}). In every variant, the calibrated static design remains the benchmark a programmable tree must beat.

\begin{table}[!htb]
\centering\small
\caption{Sensitivity of the headline quantities (maxima over the flux grid of the ensemble medians; \texttt{results/summary.json}). $G_{\mathrm{D}/\mathrm{B}^\ast}$: gain of D over the flux-aware uncalibrated fan-out.}
\label{tab:sens}
\setlength\tabcolsep{4pt}
\begin{tabular}{@{}lrrrrrr@{}}
\toprule
Variant & arrays & $\max G_\mathrm{C}$ (dB) & $\max G_\mathrm{D}$ (dB) & $\max G_{\mathrm{D}/\mathrm{B}^\ast}$ (dB) & $\max\Delta\mathrm{IL}_\mathrm{be}$ vs A & vs B$^\ast$ (dB) \\
\midrule
Nominal ($D_0=\SI{e5}{s^{-1}}$) & 64 & 15.88 & 13.18 & 1.52 & 7.97 & 2.09 \\
Second seed & 32 & 16.16 & 13.47 & 1.79 & 8.12 & 2.23 \\
Cooled ($D_0=\SI{e3}{s^{-1}}$) & 64 & 9.37 & 7.63 & $-0.13$ & 6.84 & 0.79 \\
Hot ($D_0=\SI{e6}{s^{-1}}$) & 32 & 16.14 & 13.39 & 2.04 & 8.02 & 2.34 \\
Uniform array & 1 & 12.12 & 9.39 & $-0.14$ & 6.03 & 0.00 \\
\bottomrule
\end{tabular}
\end{table}

\paragraph{Optimizer accuracy.} The allocation optimizer is a structured multistart local method, and its checks (Section~\ref{sec:pipeline}) are not a global certificate. Break-even widths come from differences between slowly growing regret curves, so small optimizer errors shift them more than any single-flux gain. They should be read as model-and-solver values with an uncertainty of a few tenths of a decade.

\section{Discussion and limitations}\label{sec:disc}
\paragraph{Design rules.} Within this model, two rules follow. First, size a passive fan-out to $k^\ast=\eta T\Phi/\sqrt{D(D+1/\tau)}$, and when the flux is known, fix the splitting set to the measured array: calibration, not programmability, captures the selection gain. Second, add MZI routing only when one chip must serve a flux range wider than about 2.5 decades, and keep its excess loss well below \SI{3}{dB} for $N=16$. At low flux, every decibel of excess loss costs about two decibels of information.

\paragraph{What the model leaves out.} The break-even shifts with temperature-dependent detector performance, thermo-optic efficiency, routing loss and heater power (15 MZIs at $\sim$\SI{25}{mW} per $\pi$ phase shift~\cite{Harris2014}); the cooled operation that lowers $D$ also changes the heater efficiency. Afterpulsing, optical and electrical crosstalk, timing jitter and any time-resolved figure of merit are omitted; with time stamps, the information available can differ from $\Gamma$. The flux is assumed stationary within $\Tobs$ and known well enough to choose the set point, and the set-point loop is assumed slow compared with $\Tobs$ but fast compared with flux changes. \Bcal{} is idealized: a fabricated static design either uses fixed MMI ratios designed from wafer-level statistics, which forgoes per-array selection, or trims on chip, which adds MZI-like loss. A fabricated static design would therefore fall short of \Bcal, and the break-even windows against it would shrink. Finally, the parameters are representative of several published devices~\cite{Vines2019,Na2024,Wang2023} and not of one measured array; the conclusions are statements about the model, to be tested against measured Ge-on-Si SPAD arrays.

\section{Conclusion}\label{sec:concl}
Splitting light over Ge-on-Si SPADs trades dead-time relief against added dark counts. The asymptotic count-based information rate turns this trade-off into closed-form design quantities: the optimal pixel load $y^\ast=\sqrt{D(D+1/\tau)}$, its efficiency $(\sqrt{1+D\tau}-\sqrt{D\tau})^2$, the optimal fan-out and the critical flux. In simulated 16-pixel arrays, ideal programmability gains up to \SI{15.9}{dB} over equal splitting, but most of that gain is available to a calibrated static design at a known flux. A thermo-optic MZI tree with \SI{1.32}{dB} excess loss pays off only when one design must serve flux windows wider than about 2.5 decades, and not within six decades once its excess loss exceeds about \SI{3.4}{dB}.

\section*{Data and code availability}
The simulation code (Python), configuration (seed 20261023), raw outputs and the scripts that produce every figure and table are available from the corresponding author. Plot data for all figures are included with the arXiv source in \texttt{data/}.

\section*{Acknowledgments}
The authors used an AI assistant (Claude, Anthropic) to write simulation and analysis code, prepare figures, and draft and revise the manuscript text. The authors reviewed the derivations, code and results and take full responsibility for the content. This work received no specific funding.

\appendix
\section{Detector-model variants}\label{app:variants}
\paragraph{Paralyzable pixel.} With $R=xe^{-x\tau}$ and $\sigma^2=xe^{-x\tau}(1-2x\tau e^{-x\tau})$, the per-pixel information is
\begin{equation}
g_\mathrm{p}(y)=\frac{y^2(1-x\tau)^2e^{-2x\tau}}{xe^{-x\tau}\,(1-2x\tau e^{-x\tau})},\qquad x=y+D,
\end{equation}
which vanishes at $x\tau=1$, where the count rate is not invertible. Table~\ref{tab:e5} compares its numerically optimal load with \eqref{eq:yopt}.

\paragraph{Gated pixel.} With detection probability per gate $p=1-e^{-(s+d)}$, the information per gate on $s$ is $(\partial p/\partial s)^2/[p(1-p)]=s^2/(e^{s+d}-1)$. Expanding $e^u-1\approx u(1+u/2)$ gives the free-running form with $\tau\to1/2$ per gate, i.e.\ $\tau\to t_g/2$ in time units, with $D$ unchanged ($d=Dt_g$ is a per-gate mean, not a rate). The predicted optimum $s^\ast\approx\sqrt{2d(1+d/2)}$ agrees with the numerical optimum within 1--\SI{7}{\percent} for $d\le0.02$ and deviates by \SI{25}{\percent} at $d=0.2$ (Table~\ref{tab:e5}).

\begin{table}[h]
\centering\small
\caption{Detector-model variants ($\tau=\SI{100}{ns}$, gate $t_g=2\tau$; \texttt{results/E5\_variants.csv}).}
\label{tab:e5}
\begin{tabular}{@{}rrrrrrrr@{}}
\toprule
$D$ (s$^{-1}$) & $D\tau$ & $y^\ast$ non-par. (s$^{-1}$) & eff. & $y^\ast$ par. (s$^{-1}$) & eff. & $s^\ast$ gated (num./pred.) & eff. gated \\
\midrule
$10^3$ & $10^{-4}$ & $1.00\times10^5$ & 0.980 & $0.99\times10^5$ & 0.980 & 0.0199 / 0.0200 & 0.980 \\
$10^4$ & $10^{-3}$ & $3.16\times10^5$ & 0.939 & $3.01\times10^5$ & 0.937 & 0.0619 / 0.0633 & 0.938 \\
$10^5$ & $10^{-2}$ & $1.00\times10^6$ & 0.819 & $0.86\times10^6$ & 0.806 & 0.187 / 0.201 & 0.813 \\
$10^6$ & $10^{-1}$ & $3.32\times10^6$ & 0.537 & $1.84\times10^6$ & 0.437 & 0.507 / 0.663 & 0.493 \\
\bottomrule
\end{tabular}
\end{table}

\bibliographystyle{IEEEtran}
\bibliography{references}
\end{document}

%% file: figs/fig_arch.tex
\begin{tikzpicture}[x=1.45mm,y=1.3mm,font=\small,>=stealth,
  wg/.style={line width=0.8pt,draw=black!75},
  sp/.style={draw,fill=orange!25,minimum width=8mm,minimum height=4.2mm,inner sep=0pt,rounded corners=0.6mm},
  ht/.style={line width=1.6pt,red!70!black},
  det/.style={draw,fill=green!18,minimum width=15mm,minimum height=5mm,inner sep=0pt},
  cnt/.style={draw,fill=blue!8,minimum width=7mm,minimum height=5mm,inner sep=0pt}]
\draw[wg] (0,0) node[left,inner sep=2pt]{$\Phi$} -- (5.2,0);
\node[sp] (s0) at (8,0) {};
\draw[ht] (6.2,1.0) -- (9.8,1.0);
\draw[wg] (10.8,0.6) .. controls (13,0.6) and (13,6) .. (15.4,6);
\draw[wg] (10.8,-0.6) .. controls (13,-0.6) and (13,-6) .. (15.4,-6);
\foreach \yy in {6,-6}{
  \node[sp] at (18.2,\yy) {};
  \draw[ht] (16.4,\yy+1.0) -- (20,\yy+1.0);
  \draw[wg] (21,\yy+0.6) .. controls (23,\yy+0.6) and (23,\yy+3) .. (25,\yy+3);
  \draw[wg] (21,\yy-0.6) .. controls (23,\yy-0.6) and (23,\yy-3) .. (25,\yy-3);}
\foreach \yy [count=\i] in {9,3,-3,-9}{
  \draw[wg] (25,\yy) -- node[above=-1pt,font=\footnotesize]{$T_{\i}s_{\i}$} (33.2,\yy);
  \node[det] at (38.5,\yy) {SPAD\,$\i$};
  \draw[->] (43.6,\yy) -- (46,\yy);
  \node[cnt] at (48.4,\yy) {$c_{\i}$};
  \draw[->] (50.8,\yy) -- (54,\yy);}
\node[draw,fill=blue!8,align=center,inner sep=2pt,minimum width=12mm,minimum height=29mm] (wm) at (58.5,0) {WM\\estimator};
\draw[->,thick] (62.7,0) -- (66,0) node[right,inner sep=1pt]{$\hat\Phi$};
\node[anchor=south] at (42,12.4) {Ge-on-Si SPADs: $\eta_i,\ D_i,\ \tau_i$};
\node[anchor=south,align=center] at (9,9.6) {$1{\times}2$ node: MMI (A, B)\\\emph{or} thermo-optic MZI (C, D)};
\draw[->,dashed,red!70!black] (58.5,-10.6) -- (58.5,-14.2) -- (18.2,-14.2) -- (18.2,-7.6);
\draw[->,dashed,red!70!black] (18.2,-14.2) -- (8,-14.2) -- (8,-1.6);
\node[anchor=north,red!70!black,inner sep=2pt] at (38.5,-14.4) {quasi-static set point from measured mean flux and pixel calibration (C, D)};
\node[anchor=north east,red!70!black,inner sep=1pt] at (6.6,-2.0) {heater};
\end{tikzpicture}

%% file: figs/fig_analytic.tex
\begin{tikzpicture}
\begin{axis}[name=pa, width=0.47\linewidth, height=52mm, xmode=log, xmin=1e3, xmax=1e11, ymin=0, ymax=1.02, xtick={1e3,1e5,1e7,1e9,1e11},
  xlabel={pre-dead-time signal rate $y$ (s$^{-1}$)}, ylabel={$g(y)/y$},
  legend style={at={(0.99,0.99)},anchor=north east}]
  \addplot[thick, blue!70!black] table[x=y,y=D3]{data/an_eff_vs_y.dat}; \addlegendentry{$D=10^3$}
  \addplot[thick, green!50!black, dashed] table[x=y,y=D4]{data/an_eff_vs_y.dat}; \addlegendentry{$10^4$}
  \addplot[very thick, black] table[x=y,y=D5]{data/an_eff_vs_y.dat}; \addlegendentry{$10^5$ (nominal)}
  \addplot[thick, red!70!black, dash dot] table[x=y,y=D6]{data/an_eff_vs_y.dat}; \addlegendentry{$10^6$\,s$^{-1}$}
  \addplot[only marks, mark=*, mark size=1.8pt, orange!85!black] table[x=ystar,y=eff]{data/an_ystar.dat}; \addlegendentry{$y^\ast$}
  \node[font=\small\bfseries, anchor=south east] at (rel axis cs:0.98,0.03) {(a)};
\end{axis}
\begin{axis}[at={(pa.east)}, anchor=west, xshift=16mm, width=0.47\linewidth, height=52mm, xmode=log, ymode=log,
  xmin=1e-5, xmax=1, ymin=1e-5, ymax=3, xlabel={$D\tau$}, ylabel={$y^\ast\tau$, \ peak efficiency},
  legend style={at={(0.98,0.03)},anchor=south east}]
  \addplot[very thick, black] table[x=u,y=ystau]{data/an_ystar_vs_u.dat}; \addlegendentry{$y^\ast\tau$ (eq.~\ref{eq:yopt})}
  \addplot[thick, blue!70!black, dashed] table[x=u,y=eff]{data/an_ystar_vs_u.dat}; \addlegendentry{efficiency (eq.~\ref{eq:yopt})}
  \addplot[only marks, mark=square, mark size=1.8pt, red!70!black] table[x=u,y=yp_tau]{data/e5.dat}; \addlegendentry{$y^\ast\tau$, paralyzable}
  \addplot[only marks, mark=triangle, mark size=2pt, red!70!black] table[x=u,y=eff_p]{data/e5.dat}; \addlegendentry{eff., paralyzable}
  \node[font=\small\bfseries, anchor=west] at (rel axis cs:0.02,0.55) {(b)};
\end{axis}
\end{tikzpicture}

%% file: figs/fig_pipeline.tex
\begin{tikzpicture}[font=\footnotesize,>=Stealth, node distance=4mm and 6mm,
  box/.style={draw,rounded corners=1mm,align=center,minimum height=10mm,text width=27mm,inner sep=2pt},
  inp/.style={box,fill=black!6}, an/.style={box,fill=orange!18}, ev/.style={box,fill=blue!8},
  mc/.style={box,fill=green!14}, outp/.style={box,fill=red!8}]
\node[inp] (cfg) {\textbf{Configuration}\\ \texttt{config.json}\\ seed 20261023};
\node[inp, below=of cfg] (gen) {\textbf{Array generator}\\ 64 arrays, $N=16$\\ $\eta_i$, $D_i$, $\tau_i$, split errors};
\node[inp, below=of gen] (opt) {\textbf{Optical model}\\ $T_i$: stages, routes,\\ 37 fluxes $10^4$--$10^{10}$\,s$^{-1}$};
\node[an, right=of cfg] (ana) {\textbf{Closed forms}\\ $R,\sigma^2,\Gamma$; $y^\ast$, $\Phi_c$\\ SymPy checks};
\node[an, right=of gen] (arch) {\textbf{Architectures}\\ A, B$_k$, B$^\ast$ (passive)\\ C, D (optimized)};
\node[an, right=of opt] (cal) {\textbf{Static benchmark}\\ B$_\mathrm{cal}$: frozen C optima,\\ $37\times37$ matrix};
\node[ev, right=of ana] (gam) {\textbf{Evaluate} $\Gamma$\\ per array, flux,\\ architecture};
\node[ev, right=of arch] (reg) {\textbf{Regret and windows}\\ worst case over\\ $w=0$--6 decades};
\node[ev, right=of cal] (be) {\textbf{Break-even}\\ IL grid 0--10\,dB;\\ loss and window};
\node[mc, right=of gam] (mc1) {\textbf{Event-level MC}\\ channel $R,\sigma^2$\\ $0.01\le x\tau\le 3$};
\node[mc, right=of reg] (mc2) {\textbf{Estimator MC}\\ WM estimator, A, C\\ 2000 trials, bootstrap};
\node[outp, right=of be] (fig) {\textbf{Figures, tables}\\ traced to \texttt{results/}\\ and \texttt{data/*.dat}};
\draw[->] (cfg) -- (gen); \draw[->] (gen) -- (opt);
\draw[->] (cfg) -- (ana); \draw[->] (gen) -- (arch); \draw[->] (opt) -- (cal);
\draw[->] (arch) -- (cal);
\draw[->] (ana) -- (gam); \draw[->] (arch) -- (reg); \draw[->] (cal) -- (be);
\draw[->] (gam) -- (reg); \draw[->] (reg) -- (be);
\draw[->] (gam) -- (mc1); \draw[->] (mc1) -- (mc2); \draw[->] (mc2) -- (fig); \draw[->] (be) -- (fig);
\end{tikzpicture}

%% file: figs/fig_performance.tex
\begin{tikzpicture}
\begin{axis}[name=pa, width=0.47\linewidth, height=54mm, xmode=log, ymode=log, xmin=1e4, xmax=1e10,
  ymin=1e-6, ymax=0.5, xtick={1e4,1e6,1e8,1e10}, minor xtick={1e5,1e7,1e9},
  xlabel={input photon rate $\Phi$ (s$^{-1}$)}, ylabel={$\Gamma/\Phi$},
  legend columns=3, legend style={at={(0.98,0.03)}, anchor=south east, fill=white, fill opacity=0.85, text opacity=1}]
  \addplot[name path=lo, draw=none, forget plot] table[x=Phi,y=Dlo]{data/fig2a.dat};
  \addplot[name path=hi, draw=none, forget plot] table[x=Phi,y=Dhi]{data/fig2a.dat};
  \addplot[orange!30, forget plot] fill between[of=lo and hi];
  \addplot[thick, black] table[x=Phi,y=A]{data/fig2a.dat}; \addlegendentry{A}
  \addplot[thick, blue!70!black, dashed] table[x=Phi,y=Bcal]{data/fig2a_bcal.dat}; \addlegendentry{B$_\mathrm{cal}$}
  \addplot[thick, green!50!black, densely dotted] table[x=Phi,y=C]{data/fig2a.dat}; \addlegendentry{C}
  \addplot[very thick, orange!85!black, dash dot] table[x=Phi,y=D]{data/fig2a.dat}; \addlegendentry{D}
  \addplot[only marks, mark=o, mark size=2pt, black] table[x=Phi,y=effA]{data/fig2a_mc.dat}; \addlegendentry{MC: A}
  \addplot[only marks, mark=triangle, mark size=2.3pt, green!40!black] table[x=Phi,y=effC]{data/fig2a_mc.dat}; \addlegendentry{MC: C}
  \node[font=\small\bfseries, anchor=north west] at (rel axis cs:0.02,0.98) {(a)};
\end{axis}
\begin{axis}[at={(pa.east)}, anchor=west, xshift=14mm, width=0.47\linewidth, height=54mm, xmode=log, xmin=1e4, xmax=1e10,
  ymin=-2, ymax=22, xtick={1e4,1e6,1e8,1e10}, minor xtick={1e5,1e7,1e9},
  xlabel={input photon rate $\Phi$ (s$^{-1}$)}, ylabel={gain over A (dB)},
  legend style={at={(0.98,0.98)}, anchor=north east}]
  \addplot[name path=clo, draw=none, forget plot] table[x=Phi,y=GClo]{data/gain_vs_flux.dat};
  \addplot[name path=chi, draw=none, forget plot] table[x=Phi,y=GChi]{data/gain_vs_flux.dat};
  \addplot[green!20, forget plot] fill between[of=clo and chi];
  \addplot[name path=dlo, draw=none, forget plot] table[x=Phi,y=GDlo]{data/gain_vs_flux.dat};
  \addplot[name path=dhi, draw=none, forget plot] table[x=Phi,y=GDhi]{data/gain_vs_flux.dat};
  \addplot[orange!30, forget plot] fill between[of=dlo and dhi];
  \addplot[thick, green!50!black] table[x=Phi,y=GC]{data/gain_vs_flux.dat}; \addlegendentry{C/A}
  \addplot[very thick, orange!85!black, dash dot] table[x=Phi,y=GD]{data/gain_vs_flux.dat}; \addlegendentry{D/A}
  \addplot[thick, blue!70!black, dashed] table[x=Phi,y=GB]{data/gain_vs_flux.dat}; \addlegendentry{B$^\ast$/A}
  \addplot[black!50, thin, forget plot] coordinates {(1e4,0) (1e10,0)};
  \node[font=\small\bfseries, anchor=north west] at (rel axis cs:0.02,0.98) {(b)};
\end{axis}
\end{tikzpicture}

%% file: figs/fig_fanout.tex
\begin{tikzpicture}
\begin{axis}[name=pa, width=0.47\linewidth, height=52mm, xmode=log, xmin=1e4, xmax=1e10, ymin=0, ymax=17,
  xtick={1e4,1e6,1e8,1e10}, minor xtick={1e5,1e7,1e9}, ytick={1,4,8,12,16},
  xlabel={input photon rate $\Phi$ (s$^{-1}$)}, ylabel={active pixels},
  legend style={at={(0.02,0.98)}, anchor=north west}]
  \addplot[name path=lo, draw=none, forget plot] table[x=Phi,y=p16]{data/nactive.dat};
  \addplot[name path=hi, draw=none, forget plot] table[x=Phi,y=p84]{data/nactive.dat};
  \addplot[green!20, forget plot] fill between[of=lo and hi];
  \addplot[thick, green!50!black, mark=*, mark size=1pt] table[x=Phi,y=med]{data/nactive.dat}; \addlegendentry{C optimum}
  \addplot[thick, black, dashed] table[x=Phi,y=kstar]{data/nactive.dat}; \addlegendentry{$k^\ast=\eta T\Phi/y^\ast$}
  \draw[red!70!black, densely dotted, thick] (axis cs:9.83e7,0) -- (axis cs:9.83e7,17);
  \node[red!70!black, anchor=south east, font=\footnotesize] at (axis cs:9.4e7,0.3) {$\Phi_c$};
  \node[font=\small\bfseries, anchor=south east] at (rel axis cs:0.98,0.02) {(a)};
\end{axis}
\begin{axis}[at={(pa.east)}, anchor=west, xshift=14mm, width=0.47\linewidth, height=52mm, xmode=log, xmin=1e4, xmax=1e10,
  ymin=0, ymax=23, xtick={1e4,1e6,1e8,1e10}, minor xtick={1e5,1e7,1e9},
  xlabel={input photon rate $\Phi$ (s$^{-1}$)}, ylabel={regret vs C (dB)},
  legend columns=3, legend style={at={(0.5,0.99)}, anchor=north}]
  \addplot[thick, blue!70!black] table[x=Phi,y=k1]{data/regret_k_vs_flux.dat}; \addlegendentry{$k=1$}
  \addplot[thick, green!50!black, dashed] table[x=Phi,y=k2]{data/regret_k_vs_flux.dat}; \addlegendentry{$k=2$}
  \addplot[thick, orange!85!black, dash dot] table[x=Phi,y=k4]{data/regret_k_vs_flux.dat}; \addlegendentry{$k=4$}
  \addplot[thick, red!70!black, densely dotted] table[x=Phi,y=k8]{data/regret_k_vs_flux.dat}; \addlegendentry{$k=8$}
  \addplot[very thick, black] table[x=Phi,y=k16]{data/regret_k_vs_flux.dat}; \addlegendentry{$k=16$ (A)}
  \node[font=\small\bfseries, anchor=south west] at (rel axis cs:0.02,0.02) {(b)};
\end{axis}
\end{tikzpicture}

%% file: figs/fig_robust.tex
\begin{tikzpicture}
\begin{axis}[name=pa, width=0.47\linewidth, height=52mm, xmin=-0.1, xmax=1.5, ymin=-1, ymax=29,
  xlabel={dark-count spread $\sigma_{\ln D}$}, ylabel={C/A gain (dB)},
  legend columns=2, legend style={at={(0.5,0.99)}, anchor=north}]
  \addplot[thick, black, mark=*, mark size=1.6pt, error bars/.cd, y dir=both, y explicit]
    table[x=sig, y=G5, y error plus expr=\thisrow{G5hi}-\thisrow{G5}, y error minus expr=\thisrow{G5}-\thisrow{G5lo}]{data/e3.dat};
  \addlegendentry{$10^5$\,s$^{-1}$}
  \addplot[thick, blue!70!black, mark=square*, mark size=1.5pt, error bars/.cd, y dir=both, y explicit]
    table[x=sig, y=G7, y error plus expr=\thisrow{G7hi}-\thisrow{G7}, y error minus expr=\thisrow{G7}-\thisrow{G7lo}]{data/e3.dat};
  \addlegendentry{$10^7$\,s$^{-1}$}
  \addplot[thick, green!50!black, mark=triangle*, mark size=1.8pt] table[x=sig, y=G9]{data/e3.dat};
  \addlegendentry{$10^9$\,s$^{-1}$}
  \addplot[thick, red!70!black, dashed] table[x=sig, y=Gclosed]{data/e3.dat};
  \addlegendentry{low-flux limit}
  \node[font=\small\bfseries, anchor=east] at (rel axis cs:0.97,0.45) {(a)};
\end{axis}
\begin{axis}[at={(pa.east)}, anchor=west, xshift=14mm, width=0.47\linewidth, height=52mm, xmode=log, xmin=1e4, xmax=1e10,
  ymin=0, ymax=14.5, xtick={1e4,1e6,1e8,1e10}, minor xtick={1e5,1e7,1e9},
  xlabel={input photon rate $\Phi$ (s$^{-1}$)}, ylabel={break-even loss vs A (dB)},
  legend columns=2, legend style={at={(0.99,0.99)}, anchor=north east}]
  \addplot[name path=lo, draw=none, forget plot] table[x=Phi,y=beAlo]{data/fig2b_nom.dat};
  \addplot[name path=hi, draw=none, forget plot] table[x=Phi,y=beAhi]{data/fig2b_nom.dat};
  \addplot[black!15, forget plot] fill between[of=lo and hi];
  \addplot[very thick, black] table[x=Phi,y=beA]{data/fig2b_nom.dat}; \addlegendentry{nominal}
  \addplot[thick, black!60, densely dotted] table[x=Phi,y=beA]{data/fig2b_seed2.dat}; \addlegendentry{second seed}
  \addplot[thick, blue!70!black, dashed] table[x=Phi,y=beA]{data/fig2b_cool.dat}; \addlegendentry{$D_0=10^3$\,s$^{-1}$}
  \addplot[thick, red!70!black, dash dot] table[x=Phi,y=beA]{data/fig2b_hot.dat}; \addlegendentry{$D_0=10^6$\,s$^{-1}$}
  \addplot[thick, green!50!black, loosely dashed] table[x=Phi,y=beA]{data/fig2b_hom.dat}; \addlegendentry{uniform array}
  \draw[red!70!black, densely dotted] (axis cs:1e4,1.32) -- (axis cs:1e10,1.32);
  \node[red!70!black, anchor=south east, font=\footnotesize] at (axis cs:8e9,1.32) {1.32\,dB (D)};
  \node[font=\small\bfseries, anchor=east] at (rel axis cs:0.97,0.45) {(b)};
\end{axis}
\end{tikzpicture}

%% file: figs/fig_windows.tex
\begin{tikzpicture}
\begin{axis}[name=ps, width=0.34\linewidth, height=54mm, xmode=log, xmin=1e4, xmax=1e10, ymin=-3, ymax=0.3,
  xtick={1e4,1e6,1e8,1e10}, minor xtick={1e5,1e7,1e9},
  xlabel={input photon rate $\Phi$ (s$^{-1}$)}, ylabel={D vs B$_\mathrm{cal}$, one flux (dB)},
  legend style={at={(0.98,0.22)}, anchor=south east}]
  \addplot[name path=lo, draw=none, forget plot] table[x=Phi,y=p16]{data/d_vs_bcal.dat};
  \addplot[name path=hi, draw=none, forget plot] table[x=Phi,y=p84]{data/d_vs_bcal.dat};
  \addplot[orange!30, forget plot] fill between[of=lo and hi];
  \addplot[very thick, orange!85!black] table[x=Phi,y=med]{data/d_vs_bcal.dat}; \addlegendentry{median}
  \addplot[thick, black, dashed] table[x=Phi,y=amax]{data/d_vs_bcal.dat}; \addlegendentry{best array}
  \addplot[black!50, thin, forget plot] coordinates {(1e4,0) (1e10,0)};
  \draw[red!70!black, densely dotted] (axis cs:1e4,-2.64) -- (axis cs:1e8,-2.64);
  \node[red!70!black, anchor=north west, font=\footnotesize] at (axis cs:1e6,-2.64) {$-2.64$\,dB};
  \draw[red!70!black, densely dotted] (axis cs:1e4,-1.32) -- (axis cs:1e8,-1.32);
  \node[red!70!black, anchor=south west, font=\footnotesize] at (axis cs:1.2e4,-1.32) {$-1.32$\,dB};
  \node[font=\small\bfseries, anchor=north west] at (rel axis cs:0.02,0.98) {(a)};
\end{axis}
\begin{axis}[name=pa, at={(ps.east)}, anchor=west, xshift=13mm, width=0.34\linewidth, height=54mm, xmin=0, xmax=6, ymin=0, ymax=12,
  xlabel={flux-window width $w$ (decades)}, ylabel={median worst-case regret (dB)}, legend columns=1,
  legend style={at={(0.02,0.98)}, anchor=north west, fill=white, fill opacity=0.85, text opacity=1}]
  \addplot[thick, blue!70!black] table[x=w,y=Bcal]{data/fig2b_regret.dat}; \addlegendentry{B$_\mathrm{cal}$}
  \addplot[thick, black, dashed] table[x=w,y=Bk]{data/fig2b_regret.dat}; \addlegendentry{B$_k$}
  \addplot[very thick, orange!85!black] table[x=w,y=D132]{data/fig2b_regret.dat}; \addlegendentry{D, 1.32\,dB}
  \addplot[thick, orange!60!black, densely dotted] table[x=w,y=D04]{data/fig2b_regret.dat}; \addlegendentry{D, 0.4\,dB}
  \addplot[thick, red!70!black, dash dot] table[x=w,y=D24]{data/fig2b_regret.dat}; \addlegendentry{D, 2.4\,dB}
  \addplot[only marks, mark=none, blue!70!black, forget plot, error bars/.cd, y dir=both, y explicit]
    table[x expr=\thisrow{w}-0.06, y=cal, y error plus expr=\thisrow{calhi}-\thisrow{cal}, y error minus expr=\thisrow{cal}-\thisrow{callo}]{data/regret_band.dat};
  \addplot[only marks, mark=none, orange!85!black, forget plot, error bars/.cd, y dir=both, y explicit]
    table[x expr=\thisrow{w}+0.06, y=d, y error plus expr=\thisrow{dhi}-\thisrow{d}, y error minus expr=\thisrow{d}-\thisrow{dlo}]{data/regret_band.dat};
  \node[font=\small\bfseries, anchor=north east] at (rel axis cs:0.98,0.98) {(b)};
\end{axis}
\begin{axis}[at={(pa.east)}, anchor=west, xshift=13mm, width=0.34\linewidth, height=54mm, xmin=0, xmax=4, ymin=0, ymax=7.4,
  xtick={0,1,2,3,4}, ytick={0,2,4,6}, unbounded coords=jump,
  xlabel={excess routing loss of D (dB)}, ylabel={break-even window (decades)},
  legend style={at={(0.02,0.98)}, anchor=north west}]
  \addplot[very thick, blue!70!black, mark=*, mark size=1.5pt] table[x=ex,y=wcal]{data/fig2b_map.dat}; \addlegendentry{vs B$_\mathrm{cal}$}
  \addplot[thick, black, dashed, mark=square, mark size=1.5pt, mark options={solid}] table[x=ex,y=wfix]{data/fig2b_map.dat}; \addlegendentry{vs best fixed $1{\times}k$}
  \draw[red!70!black, densely dotted, thick] (axis cs:1.32,0) -- (axis cs:1.32,5.4);
  \node[red!70!black, anchor=west, font=\footnotesize] at (axis cs:1.36,0.45) {1.32\,dB};
  \node[anchor=west, font=\footnotesize] at (axis cs:0.1,4.4) {D pays};
  \node[anchor=south west, font=\footnotesize] at (axis cs:2.5,0.35) {static wins};
  \node[font=\small\bfseries, anchor=north east] at (rel axis cs:0.98,0.98) {(c)};
\end{axis}
\end{tikzpicture}

%% file: figs/fig_mc.tex
\begin{tikzpicture}
\begin{axis}[name=pa, width=0.355\linewidth, height=50mm, xmode=log, xmin=5e-3, xmax=6, ymin=0.86, ymax=1.18,
  xlabel={$x\tau$}, ylabel={variance ratio MC/analytic},
  legend style={at={(0.03,0.97)}, anchor=north west}]
  \addplot[only marks, mark=*, mark size=1.6pt, black, error bars/.cd, y dir=both, y explicit]
    table[x=xtau, y=var, y error=var_se]{data/e4a_np.dat}; \addlegendentry{non-par.}
  \addplot[only marks, mark=square, mark size=1.6pt, red!70!black, error bars/.cd, y dir=both, y explicit]
    table[x expr=\thisrow{xtau}*1.2, y=var, y error=var_se]{data/e4a_p.dat}; \addlegendentry{paralyzable}
  \addplot[black!50, thin, forget plot] coordinates {(5e-3,1) (6,1)};
  \node[font=\small\bfseries, anchor=north east] at (rel axis cs:0.97,0.97) {(a)};
\end{axis}
\begin{axis}[name=pb, at={(pa.east)}, anchor=west, xshift=11mm, width=0.355\linewidth, height=50mm, xmode=log,
  xmin=3e5, xmax=3e9, ymin=0.85, ymax=1.22, xtick={1e6,1e7,1e8,1e9},
  xlabel={$\Phi$ (s$^{-1}$)}, ylabel={$\mathrm{var}(\hat\Phi)\,\Gamma\Tobs/\Phi^2$},
  legend style={at={(0.03,0.97)}, anchor=north west}]
  \addplot[only marks, mark=o, mark size=1.8pt, black, error bars/.cd, y dir=both, y explicit]
    table[x=PhiA, y=rA, y error plus=rAp, y error minus=rAm]{data/mc.dat}; \addlegendentry{A}
  \addplot[only marks, mark=triangle, mark size=2pt, green!40!black, error bars/.cd, y dir=both, y explicit]
    table[x=PhiC, y=rC, y error plus=rCp, y error minus=rCm]{data/mc.dat}; \addlegendentry{C}
  \addplot[black!50, thin, forget plot] coordinates {(3e5,1) (3e9,1)};
  \node[font=\small\bfseries, anchor=north east] at (rel axis cs:0.97,0.97) {(b)};
\end{axis}
\begin{axis}[at={(pb.east)}, anchor=west, xshift=11mm, width=0.355\linewidth, height=50mm, xmode=log,
  xmin=3e5, xmax=3e9, ymin=-1, ymax=11, xtick={1e6,1e7,1e8,1e9},
  xlabel={$\Phi$ (s$^{-1}$)}, ylabel={C/A gain (dB)},
  legend style={at={(0.97,0.97)}, anchor=north east}]
  \addplot[only marks, mark=*, mark size=1.6pt, black, error bars/.cd, y dir=both, y explicit]
    table[x=Phi, y=G, y error plus=Gp, y error minus=Gm]{data/mc.dat}; \addlegendentry{MC}
  \addplot[only marks, mark=x, mark size=3pt, very thick, red!70!black] table[x=Phi, y=Gpred]{data/mc.dat}; \addlegendentry{predicted}
  \node[font=\small\bfseries, anchor=south west] at (rel axis cs:0.03,0.03) {(c)};
\end{axis}
\end{tikzpicture}